\documentclass[a4paper,11pt]{article}
\usepackage{pos}
\usepackage{verbatim,ulem,amsmath,scalefnt}
\usepackage[capitalize]{cleveref}
\usepackage{slashed}
\newcommand{\lmut}{l_{\mu t}}

\newcounter{notecount}

\newcommand{\ep}{\epsilon}
\newcommand{\bare}{\text{\abbrev{B}}}
\newcommand{\abbrev}[1]{{\scalefont{.9}#1}}
\newcommand{\msbar}{\ensuremath{\overline{\mbox{\abbrev{MS}}}}}

\newcommand{\nklo}[1]{\abbrev{N$^{#1}$LO}}
\newcommand{\nnlo}{\abbrev{NNLO}}
\newcommand{\nlo}{\abbrev{NLO}}
\newcommand{\qcd}{\abbrev{QCD}}
\newcommand{\lqcd}{\abbrev{LQCD}}
\newcommand{\edm}{\abbrev{EDM}}
\newcommand{\sm}{\abbrev{SM}}
\newcommand{\bsm}{\abbrev{BSM}}
\newcommand{\cp}{\abbrev{CP}}
\newcommand{\qcedm}{q\abbrev{CEDM}}

\title{Two-loop matching of the chromo-magnetic dipole operator
  with the gradient flow}
\ShortTitle{Chromo-magnetic dipole operator in gradient flow}

\author[a]{Janosch Borgulat}
\author*[a]{Robert Harlander}
\author[b]{Matthew D.~Rizik}
\author[b]{Andrea Shindler}

\affiliation[a]{TTK, RWTH Aachen University, D-52056 Aachen, Germany}

\affiliation[b]{FRIB \& Physics Department,
Michigan State University, East Lansing, MI 48824, USA}

\emailAdd{janosch.borgulat@rwth-aachen.de}
\emailAdd{harlander@physik.rwth-aachen.de}
\emailAdd{rizik@nscl.msu.edu}
\emailAdd{shindler@frib.msu.edu}

\abstract{The chromo-magnetic dipole operator is expressed in terms of 
  operators at finite flow time in the gradient-flow formalism. The matching
  coefficients are evaluated through next-to-next-to-leading order \qcd.}

\FullConference{%
 The 39th International Symposium on Lattice Field Theory, LATTICE2022}

\begin{document}
\maketitle

%- }}}
%- {{{ section{Introduction}

\section{Introduction}
\label{sec:intro}

Beyond the Standard Model (\bsm) effects on observables measured at hadronic
scales of a few GeV are encoded in higher dimensional ($D>4$) operators.  The
relevant degrees of freedom are the same as the ones of \qcd, quarks and
gluons, and \abbrev{QED} effects should also be included if the precision
requires it.  In most cases, the calculation of hadronic matrix elements using
lattice \qcd\ (\lqcd) is challenging because of the complicated
renormalization pattern.

A long-standing unresolved question in modern physics is the measured
matter-antimatter asymmetry in the Universe~\cite{ParticleDataGroup:2022pth},
that is orders of magnitudes larger than the Standard Model (\sm)
prediction~\cite{Gavela:1993ts,Huet:1994jb}.  Sakharov
conditions~\cite{Sakharov:1967dj}, among other constraints, dictate that any
interaction responsible for the baryon asymmetry should violate \cp-symmetry.
The amount of \cp-violation in the \sm\ (from the \abbrev{CKM} matrix) is
wildly insufficient to explain the measured baryon asymmetry, thus providing a
very strong argument to search for new sources of \cp-violation.

The electric dipole moment (\edm) of the neutron provides a system sensitive
to \cp-violating \bsm\ sources and free of any background for many order of
magnitude.  Many different higher dimensional operators contribute to the
neutron \edm\ and in particular the so-called $D=5$ quark-chromo
\edm\ (\qcedm) provides an important contribution (see
Ref.~\cite{Shindler:2021bcx} for a review on \lqcd\ calculations on the
neutron \edm).  The \qcedm\ has a complicated renormalization pattern on the
lattice~\cite{Bhattacharya:2015rsa}, which has prevented the calculation of
the \qcedm\ contribution to the neutron \edm.

In a set of publications
~\cite{Shindler:2014oha,Dragos:2017wms,Dragos:2018uzd,Kim:2018rce,
  Rizik:2020naq,Mereghetti:2021nkt}, it was proposed to use the gradient
flow~\cite{Luscher:2010iy,Luscher:2011bx,Luscher:2013cpa}, to solve the
renormalization of the \qcedm\ as well as all the other higher dimensional
operators contributing to the neutron \edm.  The method proposed is based on
the short flow-time expansion of the higher dimensional operators and it needs
non-perturbative \lqcd\ computations combined with calculations in
perturbative \qcd. Perturbative \qcd\ is needed to match the \lqcd\ results
obtained at finite flow time $t$, with the renormalized matrix elements at
$t=0$.  In Refs.~\cite{Rizik:2020naq,Mereghetti:2021nkt}, a 1-loop calculation
of the matching coefficients of the \qcedm\ and the corresponding \cp-even
operator, the chromo-magnetic (CM) operator was presented.

Using the methods and tools of Ref.\,\cite{Artz:2019bpr}, it is possible to
extend this calculation to the 2-loop level, as it has been done in similar
applications including the energy-momentum tensor\,\cite{Harlander:2018zpi},
the hadronic vacuum polarization~\cite{Harlander:2020duo}, and the effective
weak Hamiltonian~\cite{Harlander:2022tgk} (see also
Ref.\,\cite{Harlander:2021esn}). In these proceedings we present this
extension to two loops for the chromo-magnetic dipole operator in the case of
massless quark. This avoids nuisances with the definition of $\gamma_5$ in a
generic $D$ dimension.  We emphasize that the calculation of the matching
coefficients on the CM operator is a phenomenologically relevant calculation
on its own. For example for strangeness changing $\Delta S = 1$ CM operators
contribute to rare kaon decays (\cp-even and \cp-odd), to the $K^0 -
\overline{K}^0$ oscillation and to the $\epsilon'/\epsilon$ ratio
parametrizing direct \cp-violation~\cite{Buras:1999da,DAmbrosio:1999exg}.

%- }}}
%- {{{ section{Operator basis}

\section{Operator basis}

Throughout this paper, we work in single-flavor massless \qcd. The general
case will be deferred to a forthcoming publication.  Working in $D=4-2\ep$
space-time dimensions in order to regularize \abbrev{UV} and \abbrev{IR}
divergences, the quark chromomagnetic dipole operator is
\begin{equation}\label{eq::fist}
  \begin{aligned}
    \mathcal{O}_{CM} &= \mu^{2\ep}
    g_\bare \bar\psi t^a\sigma_{\mu\nu}\psi F^a_{\mu\nu}\,,
  \end{aligned}
\end{equation}
where we factored out one power of the bare strong coupling $g_\bare$, and
introduced the 't~Hooft mass $\mu$ in order to compensate for the non-integer
mass dimension of the operator. $\psi$ is the bare quark field, $t^a$ the
generators of SU(3) in the fundamental representation, and the field strength
tensor is given by
\begin{equation}\label{eq::garm}
  \begin{aligned}
   F^a_{\mu\nu} = \partial_\mu A_\nu^a - \partial_\nu A_\mu^a +
   g_B f^{abc}A_\mu^bA_\nu^c\,,
  \end{aligned}
\end{equation}
with the bare gluon field $A_\mu$ and the SU(3) structure constants $f^{abc}$.
The goal of this paper is to express this operator in terms of its flowed
counter part, defined through
\begin{equation}\label{eq::fista}
  \begin{aligned}
    \tilde{\mathcal{O}}_{CM}(t) &= g \bar\chi(t)
    t^a\sigma_{\mu\nu}\chi(t) G^a_{\mu\nu}(t)\,,
  \end{aligned}
\end{equation}
with $g$ the $\msbar$-renormalized strong coupling.
The flowed quark and gluon fields are defined through the flow
equations
\begin{equation}\label{eq::fane}
  \begin{aligned}
\partial_t \chi(x,t) = \Delta \chi(x,t)\,, \qquad
 \partial_t B_\mu(x,t) = D_\nu G_{\nu\mu}(x,t)\,,
  \end{aligned}
\end{equation}
where $\Delta = D_\mu D_\mu$ contains derivative acting on the fundamental
representation, while the derivative acting on the field tensor reads $D_\mu =
\partial_\mu + \left[ B_\mu, \cdot\right]$.  The boundary conditions satisfied
by the fields are
\begin{equation}\label{eq::icky}
  \begin{aligned}
    \chi(t=0)=\psi\,,\qquad B^a_{\mu}(t=0) = A^a_{\mu}\,.
  \end{aligned}
\end{equation}
Taking the limit $t\to 0$ of $\tilde{\mathcal{O}}_{CM}(t)$ leads to an
operator-product expansion in terms of regular-\qcd\ operators with
$t$-dependent coefficients.

Up to leading order in $t$, it involves $\mathcal{O}_{CM}$,
the dimension-3 operator
\begin{equation}\label{eq::gaur}
  \begin{aligned}
    \mathcal{O}_{S} &=i \mu^{2\ep}\bar\psi\psi\,,
  \end{aligned}
\end{equation}
plus operators that vanish in physical matrix elements due to gauge invariance
or equations of motion. Even though the latter can be neglected in the final
result, they will play a role in our calculation as outlined in the next
section.

In order to express $\mathcal{O}_{CM}$ in terms of flowed operators, we
therefore need to introduce also the flowed counter part
$\tilde{\mathcal{O}}_S(t)$ of $\mathcal{O}_S$, which allows us to establish a
one-to-one relation between regular and flowed operators, given by the bare
matching matrix $\zeta^\bare(t)$:
\begin{equation}\label{eq::brit}
  \begin{aligned}
    \left(
    \begin{matrix}
      \tilde{\mathcal{O}}_{CM}(t)\\
      \tilde{\mathcal{O}}_{S}(t)
    \end{matrix}
    \right)=\zeta^\bare(t)\,
    \left(
    \begin{matrix}
      \mathcal{O}_{CM}\\
      \mathcal{O}_{S}
    \end{matrix}
    \right) + \cdots\,,
  \end{aligned}
\end{equation}
where
\begin{equation}\label{eq::duad}
  \begin{aligned}
       \tilde{ \mathcal{O}}_{S}(t) &= i\bar\chi(t)\chi(t)\,,
  \end{aligned}
\end{equation}
and the ellipsis denotes terms that vanish as $t\to 0$. Since the flowed
operators lead to \abbrev{UV} finite Green's functions,
while those of the regular operators are divergent in general, the
$\zeta^\bare(t)$ will contain \abbrev{UV} singularities in general. We can,
however, define a renormalized matching matrix as
\begin{equation}\label{eq::icao}
  \begin{aligned}
    \zeta(t) = \zeta^\bare(t)Z^{-1}
    = \left(
    \begin{matrix}
      c_{CM}(t) & c_S(t)/t\\
      0+\cdots & s_S(t)
    \end{matrix}
    \right)\,,
  \end{aligned}
\end{equation}
where again the ellipsis denotes terms that vanish as $t\to 0$.  The operator
renormalization matrix $Z$ is diagonal in our case, because operators of
different mass dimension do not mix in the \msbar\ scheme:
\begin{equation}\label{eq::idea}
  \begin{aligned}
    Z = \left(
    \begin{matrix}
      Z_{CM} & 0\\
      0 & Z_m
    \end{matrix}
    \right)\,.
  \end{aligned}
\end{equation}
$Z_m$ is the quark mass renormalization constant, and $Z_{CM}$ can be
extracted through \nklo{3} from Ref.\,\cite{Gorbahn:2005sa}.

%- }}}
%- {{{ section{Determination of the matching matrix}

\section{Determination of the matching matrix}

The elements of the matching matrix $\zeta^\bare$ can be determined with the
help of the method of projectors~\cite{Gorishnii:1983su}. The projections
consist of certain derivatives of suitable Green's functions w.r.t.\ external
momenta, collectively denoted by $p$ in the following, such that\footnote{The
general case also includes derivatives w.r.t.\ masses.}
\begin{equation}\label{eq::afar}
  \begin{aligned}
    P_i[\mathcal{O}_j] \equiv \mathcal{P}_i(\partial_p)\langle
    i|\Gamma_i\mathcal{O}_j|0\rangle\bigg|_{p=0} = \delta_{ij}\,,
  \end{aligned}
\end{equation}
where $\Gamma_i$ is a matrix in Dirac space, $\mathcal{P}_i(x)$ is polynomial
in $x$, and $\langle i|$ is a state in (adjoint) Fock space.  It is important
that the derivatives $\partial_p$ and the nullification of $p$ are understood
to be taken before any loop integration is carried out. This means that it is
sufficient to satisfy \cref{eq::afar} at tree-level, because all loop
contributions are scaleless and vanish in dimensional regularization.

Using these projections, one thus directly obtains
\begin{equation}\label{eq::fess}
  \begin{aligned}
    \zeta^\bare_{ij}(t) = P_j[\tilde{\mathcal{O}}_i(t)]\,.
  \end{aligned}
\end{equation}
Since the Green's functions considered in \cref{eq::afar} do not represent
physical matrix elements, one needs to take into account operators which
vanish by equations-of-motion in the determination of the projectors.

\begin{figure}
  \begin{center}
    \begin{tabular}{ccc}
      \raisebox{0em}{%
        \mbox{\includegraphics[%
              viewport=150 500 380 705,%
              clip,
            width=.2\textwidth]%
                          {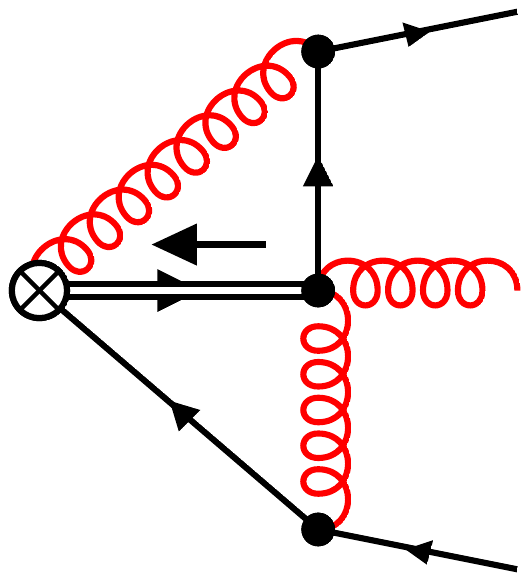}}
      } &
      \raisebox{0em}{%
          \includegraphics[%
            viewport=150 500 380 705,%
            clip,
            width=.2\textwidth]%
                          {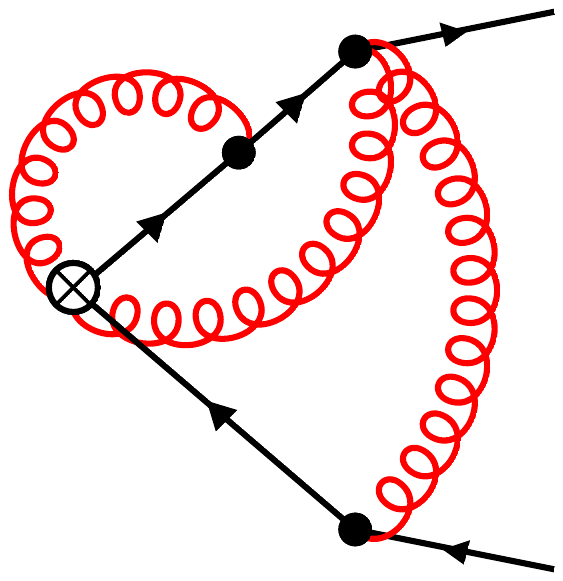}
      } &
      \raisebox{0em}{%
          \includegraphics[%
            viewport=150 500 380 705,%
            clip,
            width=.2\textwidth]%
                          {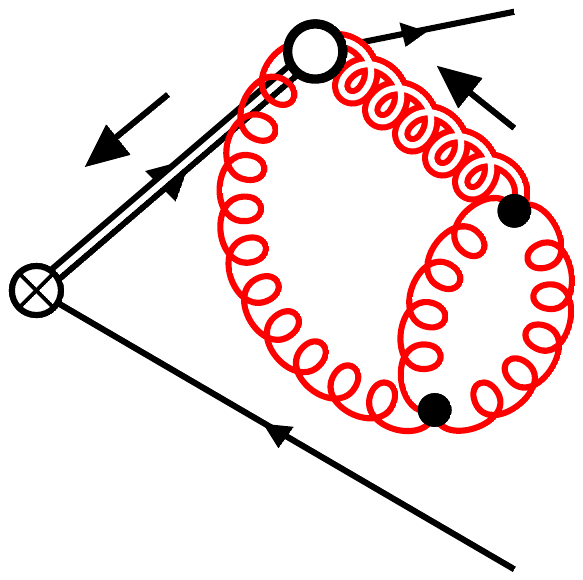}
      } \\[-1em]
      (a) & (b) & (c)
    \end{tabular}
    \parbox{.9\textwidth}{
      \caption[]{\label{fig::diagrams}\sloppy Examplary (non-vanishing)
        diagrams for the calculation of $c^\bare_{CM}(t)$, $c^\bare_{S}(t)$,
        and $s_S^\bare(t)$, (a)--(c). Double lines denote ``flow-lines'', the
        direction of the flow is indicated by the arrow next to the line. The
        vertex on the left is due to the flowed operator. Vertices with white
        filling are at finite flow time. See Ref.\,\cite{Artz:2019bpr} for
        more details. Diagrams produced using
        \texttt{FeynGame}\,\cite{Harlander:2020cyh}.  }}
  \end{center}
\end{figure}

For example, it is suggestive to define the following projector for
$\mathcal{O}_{CM}$:
\begin{equation}\label{eq::kina}
  \begin{aligned}
    P'_{CM}[X] &= -i\frac{t^a_{ij}}
    {16\,g_\bare D(D-1)}\frac{\partial}{\partial q_\rho}
    \langle ija\mu,p,q|\sigma^{\mu\rho} X|0\rangle\bigg|_{q=p=0}\,,
  \end{aligned}
\end{equation}
where $\langle ija\mu,p,q|$ contains a quark and an antiquark of color
$i$ and $j$ and momenta $p$ and $-(p+q)$, as well as a gluon of color $a$,
Lorentz index $\mu$, and momentum $q$.  Indeed, this gives
$P'_{CM}[\mathcal{O}_{CM}]=1$ and $P'_{CM}[\mathcal{O}_{S}]=0$. However, one
easily shows that $P'_{CM}[\mathcal{O}_{D^2}] = -i/2$, where
\begin{equation}\label{eq::biro}
  \begin{aligned}
    \mathcal{O}_{D^2} &= \mu^{2\ep}\bar\psi\slashed{D}^2\psi\,,
  \end{aligned}
\end{equation}
In order to subtract this contribution in the proper definition of the
projector onto $\mathcal{O}_{CM}$, we define a projector onto
$\mathcal{O}_{D^2}$ as
\begin{equation}\label{eq::jany}
  \begin{aligned}
    P_{D^2}[X] &=
    i\frac{t^a_{ij}}{2Dg_\bare}\frac{\partial}{\partial q_\mu}
    \langle ija\mu,p,q|X|0\rangle\bigg|_{q=p=0}\,,
  \end{aligned}
\end{equation}
which leads to the actual projector onto $\mathcal{O}_{CM}$:
\begin{equation}\label{eq::jest}
  \begin{aligned}
    P_{CM} = P'_{CM} + \frac{i}{2}P_{D^2}\,.
  \end{aligned}
\end{equation}
The projector onto $\mathcal{O}_S$ is simply
\begin{equation}\label{eq::cowl}
  \begin{aligned}
    P_{S}[X] &=
    -i\frac{\delta_{ij}}{12}
    \langle ij,p|X|0\rangle\bigg|_{p=0}\,,
  \end{aligned}
\end{equation}
where $\langle ij,p|$ contains a quark-antiquark pair of color $ij$ and
momenta $p$ and $-p$.  In total, this amounts to 3375 Feynman diagrams for
$c^\bare_{CM}$ at 2-loop level (45 at 1-loop level), 226 (5) for $c^\bare_S$,
and 383 (10) for $s^\bare_S$. One examplary 2-loop diagram for each of these
coefficients is shown in Fig.\,\ref{fig::diagrams}.  We compute them using the
setup described in Ref.\,\cite{Artz:2019bpr}.  After performing the Dirac
traces and setting the external momenta to zero, they lead to integrals of the
form
\begin{equation}\label{eq::cent}
  \begin{aligned}
    I(\mathbf{c},\mathbf{a},\mathbf{b}) =
    \int_0^1 d \mathbf{u}\, \mathbf{u}^\mathbf{c}
    \int \frac{d^D \mathbf{p}}{(2\pi)^{lD}}
    \frac{\exp\left[-t \mathbf{a}(\mathbf{u})\cdot\mathbf{D}(\mathbf{p})\right]}
    {\mathbf{D}^\mathbf{b}(\mathbf{p})}
  \end{aligned}
\end{equation}
where the $\mathbf{D}=(D_1,\ldots,D_k)$ are quadratic polynomials of the
$\mathbf{p}=(p_1,\ldots,p_l)$, and the $\mathbf{a}=(a_1,\ldots,a_k)$ are
polynomials of the (dimensionless) flow-time variables
$\mathbf{u}=(u_1,\ldots,u_f)$. Furthermore,
$\mathbf{u}^\mathbf{c}=u_1^{c_1}\cdots u_f^{c_f}$ and
$\mathbf{D}^\mathbf{b}=D_1^{b_1}\cdots D_k^{b_k}$, with integer $c_i$ and
$b_k$. At one-loop level, the parameters take the values $l=1$, $0\leq f\leq
2$, and $k=1$, while at two-loop level, it is $l=2$, $0\leq f\leq 4$, and
$k=3$.

%- {{{ fig::diagrams

%
%

%- }}}

Using integration-by-parts\,\cite{Chetyrkin:1981qh,Tkachov:1981wb},
including its extension to flow-time integrals\,\cite{Artz:2019bpr}, it is
possible to reduce all one-loop integrals to the single master integral
\begin{equation}\label{eq::iris}
  \begin{aligned}
    I(\{\},\{2\},\{0\}) = \int \frac{d^Dp}{(2\pi)^D}\,e^{-2tp^2}
    = \frac{1}{(8\pi t)^{D/2}}\,.
  \end{aligned}
\end{equation}
At two-loop level, we find four master integrals:
\begin{equation}\label{eq::baor}
  \begin{aligned}
    I(\{\},\{2,2,0\},\{0,0,0\}) &= \frac{1}{(8\pi t)^D}\,,\\
    I(\{\},\{2,0,0\},\{1,1,0\}) &= \frac{t^2}{(8\pi t)^D}
    \left(\frac{4}{\ep} + 4 + O(\ep)\right)\,,\\
    I(\{0\},\{2-u_1,u_1,u_1\},\{0,0,0\}) &= \frac{1}{(8\pi t)^D}
    \left( \frac{1}{2\ep}-\frac{7}{6} + \ln2 - \frac{1}{2} \ln3
    + O(\ep)\right)\,,\\
    I(\{0\},\{1-u_1,1+u_1,1+u_1\},\{0,0,0\}) &= \frac{1}{(8\pi t)^D}
    \left(\frac{2}{3} + \frac{1}{2}\ln3 + O(\ep)\right)\,.
  \end{aligned}
\end{equation}

%- }}}
%- {{{ section{Result}

\section{Result}
After renormalization according to \cref{eq::icao}, the dimensionless
coefficients of the matching matrix $\zeta(t)$ through \nnlo\ \qcd\ read, in
single-flavor \qcd,
\begin{equation}\label{eq::hock}
  \begin{aligned}
    c_{CM} &= 1 + \frac{\alpha_s}{\pi}\left(
      -4.0228 + 0.16667\,\lmut
      \right)
    + \left(\frac{\alpha_s}{\pi}\right)^2\bigg(
      -11.61 - 10.15\,\lmut
      + 0.2292\,\lmut^2
      \bigg)\,,\\
    c_{S} &= -2\frac{\alpha_s}{\pi}
    + \left(\frac{\alpha_s}{\pi}\right)^2
    \left(
    -6.136 - 3.167\,\lmut
    \right)\,,\\
    s_{S} &= 1 + \frac{\alpha_s}{\pi}\left(-2.6895 - \lmut\right)
    + \left(\frac{\alpha_s}{\pi}\right)^2
    \bigg(
      -4.546 - 8.328\,\lmut - 0.7917\,\lmut^2
      \bigg)\,,
  \end{aligned}
\end{equation}
where $\alpha_s=g^2/(4\pi)$, and $l_{\mu t}=\text{log}(2\mu^2t)+\gamma_E$,
with $\gamma_E=0.577216\ldots$ the Euler-Mascheroni constant. For the sake of
brevity, we have inserted SU(3) color factors in this expression, and
expressed all coefficients as floating point numbers. Through \nlo, the
coefficients of $c_{CM}$ and $c_S$ agree with
Ref.\,\cite{Mereghetti:2021nkt},\footnote{Note the different normalization of
the operators and matching coefficients though.} all other results are new.

%- }}}
%- {{{ section{Summary}

\section{Summary}

The renormalization of the chromo-magnetic dipole operator on the lattice is
particularly challenging because of the mixing with operators of the same and
lower dimensions.  The gradient flow provides a tool to resolve this
challenge, because it allows a matching of the calculation of matrix elements
of flowed operators with the renormalized physical ones at vanishing flow
time.  While the matching with lower dimensional operators will eventually
have to be performed non-perturbatively, a perturbative calculation provides a
strong guidance when analyzing lattice \qcd\ data.  We have presented a 2-loop
calculation of the matching coefficients for massless quark. The extension to
include operators with non-vanishing mass is in progress.  Beside a direct
application to kaon physics, these results pave the way to determination of
renormalized \cp-odd quark-chromo electric dipole moment operator, a very
important contribution from beyond-the-standard-model physics to the neutron
electric dipole moment.

%- }}}
%- {{{ section{Acknowledgements}

\section*{Acknowledgements}

\noindent
JB and RH are supported by Deutsche Forschungsgemeinschaft (grant
HA~2990/10-1).  MDR and AS acknowledge funding support under the National
Science Foundation grant PHY-1913287 and PHY-2209185.

%- }}}

\bibliographystyle{JHEP}
\bibliography{INSPIRE-CiteAll.bib}
\end{document}